%% file: main.tex
\documentclass[aps,prl,superscriptaddress,floatfix,onecolumn, notitlepage, preprint,nobibnotes]{revtex4-2}
\usepackage{amsmath,amssymb,amsfonts}
\usepackage{hyperref}
\usepackage{graphicx}
\usepackage{ulem}
\usepackage[table,x11names]{xcolor}
\usepackage{gensymb}
\usepackage{bm}
\usepackage{physics}
\usepackage{colortbl}
\usepackage{booktabs}

\newcommand{\myrowcolour}{\rowcolor[gray]{0.925}}
\makeatletter

    \def\CT@@do@color{%
      \global\let\CT@do@color\relax
            \@tempdima\wd\z@
            \advance\@tempdima\@tempdimb
            \advance\@tempdima\@tempdimc
    \advance\@tempdimb\tabcolsep
    \advance\@tempdimc\tabcolsep
    \advance\@tempdima2\tabcolsep
            \kern-\@tempdimb
            \leaders\vrule
                    \hskip\@tempdima\@plus  1fill
            \kern-\@tempdimc
            \hskip-\wd\z@ \@plus -1fill }
    \makeatother

\makeatletter
\def\blfootnote{\xdef\@thefnmark{}\@footnotetext}
\makeatother

\makeatletter
\newcommand{\fmarki}{*}
\newcommand{\fmarkii}{\ensuremath{\dagger}}
\newcommand{\fmarkiii}{\ensuremath{\ddagger}}
                
\def\@fnsymbol#1{{\ifcase#1\or \fmarki\or \fmarkii\or \fmarkiii \else\@ctrerr\fi}}
\makeatother
\renewcommand{\fmarki}{\ensuremath{\dagger}}
\renewcommand{\fmarkii}{\ensuremath{\ddagger}}
\renewcommand{\fmarkiii}{\ensuremath{\mathsection}}

\include{main_figure_definitions}

\begin{document}

\def \hf {HfTe$_5$}
\def\hfd{HfTe$_{5-\delta}$}
\def\zr{ZrTe$_5$}
\def\eb{$E_\mathrm{B}$}  
\def\abandchi{$\alpha$, $\beta$, and $\chi_i$}

\title{On the effects of strain, defects, and interactions on the topological properties of  \texorpdfstring{\hf}{HfTe5}}

\author{Na Hyun Jo\textsuperscript{*}\blfootnote{\hspace{-1em}\textsuperscript{*} These authors contributed equally}}
\email[]{nhjo@umich.edu}
\affiliation{Advanced Light Source, Lawrence Berkeley National Laboratory, Berkeley, CA 94720, USA}
\affiliation{Department of Physics, University of Michigan, Ann Arbor, Michigan 48109, USA}

\author{Omar A. Ashour\textsuperscript{*}}          
\affiliation{Department of Physics, University of California, Berkeley, California 94720, USA}
\affiliation{Materials Science Division, Lawrence Berkeley National Laboratory, Berkeley, CA 94720, USA}
\affiliation{Molecular Foundry, Lawrence Berkeley National Laboratory, Berkeley, CA, 94720, USA}

\author{Zhixue Shu}
\affiliation{Department of Physics, University of Arizona, Tucson, AZ 85721, USA}

\author{Chris Jozwiak}
\affiliation{Advanced Light Source, Lawrence Berkeley National Laboratory, Berkeley, CA 94720, USA}

\author{Aaron Bostwick}
\affiliation{Advanced Light Source, Lawrence Berkeley National Laboratory, Berkeley, CA 94720, USA}

\author{Sae Hee Ryu}
\affiliation{Advanced Light Source, Lawrence Berkeley National Laboratory, Berkeley, CA 94720, USA}

\author{Kai Sun}
\affiliation{Department of Physics, University of Michigan, Ann Arbor, Michigan 48109, USA}

\author{Tai Kong}
\affiliation{Department of Physics, University of Arizona, Tucson, AZ 85721, USA}

\author{Sinéad M. Griffin}
\email[]{sgriffin@lbl.gov}
\affiliation{Materials Science Division, Lawrence Berkeley National Laboratory, Berkeley, CA 94720, USA}
\affiliation{Molecular Foundry, Lawrence Berkeley National Laboratory, Berkeley, CA, 94720, USA}

\author{Eli Rotenberg}
\email[]{erotenberg@lbl.gov}
\affiliation{Advanced Light Source, Lawrence Berkeley National Laboratory, Berkeley, CA 94720, USA}

\date{\today}
\maketitle

\def\tmax{$T_\mathrm{max}$} 

\def\imskw{Im$\Sigma(\omega, k)$}
\def\dk{$\Delta k_{\mathrm{bulk}}(\omega)$}

\textbf{Topological insulators are characterized by spin-momentum-locked massless surface states which are robust under various perturbations. Manipulating such surface states is a topic of vigorous research, as a possible route for the realization of emergent many-body physics in topological systems. Thus far, time-reversal symmetry breaking via Coulomb and magnetic perturbations has been a dominant approach for the tuning of topological states. However, the effect of the structural degrees of freedom on quasi-particle dynamics in topological materials remains elusive. In this work, we demonstrate a transition in \hf\ between distinct topological phases as a function of either Te vacancy concentration or applied strain; these phases are characterized theoretically as a transition from strong to weak topological insulator and experimentally by a transition from sharp surface states and Dirac crossing to a Fermi-liquid-like quasiparticle state in which these surface-localized features are heavily suppressed. Although vacancies can result in various consequences such as scattering, doping, and structural distortions, we show that changes in the lattice constants play the foremost role in determining the electronic structure, self-energy, and topological states of \hf.  Our results demonstrate the possibility of using both defect chemistry and strain as control parameters for topological phase transitions and associated many-body physics.}

\figone

Unlike traditional phase transitions, which are described by symmetry breaking and the onset of an order parameter, quantum spin Hall (QSH) phases arise only for electronic states of particular topologies. The change in the electronic topology between a bulk material and the vacuum gives rise to protected surface states. Ever since topological insulators were realized in three-dimensional bulk materials, tuning their associated QSH edge states has been of great interest. Although such gapless surface states are robust with respect to various perturbations, time-reversal-symmetry (TRS) breaking through Coulomb and magnetic perturbations is known to open a gap, for example in the presence of randomly distributed Fe atoms \cite{chen2010massive}. Recent studies further imply that exotic states, such as axion insulators, can arise in the presence of ordered magnetic atoms with certain crystalline symmetries \cite{Zhang2019prl,li2019intrinsic,otrokov2019,Jo2020, Frey_et_al:2020}. While controlling topological states with the spin degree of freedom has been thoroughly explored, other degrees of freedom, including lattice, charge, and orbital orders, are less prevalent in the literature. 

After the discovery of topological materials in the 2010s, \hf\ and \zr\ gained attention due to the range in their reported, and often conflicting, topological properties. Experimentally, \hf\ and \zr\ were variously reported to be Dirac semimetals, weak topological insulators (WTIs), or strong topological insulators (STIs)\cite{li2016chiral,Chen2015,Chen20152,Zheng2016,Shen2017,Wu2016prx,Li2016prl,Manzoni2015,zhang2017nc,Luca2016,Manzoni2016,liu2018, Nair_et_al:2018}. Discrepancies exist between reports not only for the reported topological properties but also regarding temperature-dependent band changes\cite{McIlroy_2004,Manzoni2015prl,zhang2017,Luca2016,Wu2016prx} and novel magnetotransport properties\cite{tang2019,liu2016,galeski2021}. These discrepancies have been attributed to either different sample defects or different measurement conditions. Considering the most recent hydrostatic pressure and uniaxial strain studies on \zr\ \cite{Zhou2016,mutch2019ev,zhang2021}, which demonstrate tuning the topological phases with strain/pressure, the lattice must also have a significant effect on these materials. Diverging claims from first-principles calculations have revealed the sensitivity of the electronic and topological properties to the structural parameters, choice of exchange-correlation functional, and often-neglected finite-temperature effects\cite{Hochberg_et_al:2018, Monserrat/Narayan:2019}. Nonetheless, these works all highlight the extreme sensitivity of \hf\ and \zr\ to external perturbations, placing them near critical points for electronic and topological phase transitions. In fact ZrTe$_5$ has been suggested as a low-mass dark matter detection target owing to its proximity to this critical point, motivating the accurate characterization and control of its topological and electronic properties\cite{Hochberg_et_al:2018}. Because the vacancy defects alter the local bonding environment as well as the local bond lengths, it remains unclear which of these factors has a more pronounced effect on the electronic structure and hence on the observed topological phase of these systems. 

In this work, we successfully synthesized single crystals of \hf\ with different vacancy densities. We studied the impact of Te vacancy concentration using a combination of electronic transport and angle-resolved photoemission spectroscopy (ARPES). Dramatic transformations in the band structure, including the appearance and disappearance of surface-localized states meeting at a Dirac crossing, were observed with an increasing number of vacancies. Crucially, we were able to reverse such changes in the band structure with applied uniaxial stress characterized by a liquid-like quasiparticle spectral function. We present density functional theory (DFT) calculations incorporating either different strains or Te vacancies, demonstrating similar bandstructure transformations with lattice constant, and supporting the existence of a topological phase transition in \hf\ with either strain or Te vacancies. This study establishes structural degrees of freedom such as strain and vacancies as promising control parameters for topology, akin to other orders in quantum materials.

\figtwo-theory-defects

We grew single crystals of \hf\ with different concentrations of Te-vacancies \cite{Lv2018Tu} by adjusting annealing temperatures after the growth (details can be found in Methods). Electrical transport measurements were performed to characterize those samples. As shown in Fig.\ \ref{fig:Fig1_1}, the temperature of maximum relative resistance, \tmax,  for sample 1 is higher ($\sim 50K$) than that of sample 2 ($\sim 20K$). Additionally, magnetoresistance (MR=100$\times(R(H)-R(0))/R(0)$)) behaviors are very different: first, sample 1 has almost one order higher MR value at 90~kOe compared to the MR of sample 2. Second, MR saturates around 20~kOe for sample 2, while it does not saturate until 90~kOe for sample 1. These electrical transport behavior reproduce the results of reference \cite{Lv2018Tu} quite well. By comparing our transport results to \cite{Lv2018Tu}, it is clear that sample 2 has fewer Te vacancies than sample 1, likely a result of its lower post-annealing temperature. The observed linear relationship between $T_{max}$ and $\delta$ \cite{Lv2018Tu} in \hfd\cite{Lv2018Tu} was used to estimate $\delta\sim$(0.066, 0.022) for samples (1,2), respectively (see Fig.\ \ref{fig:supFigResistivity}).

We first consider the influence of Te vacancies on our structural and electronic properties via DFT calculations for \hf\ in the $Cmcm$ structure. The conventional unit cell consists of four Hf-Te zig-zag  chains along the unit cell $a$ direction, as indicated in Fig. \ref{fig:theory-defects}(a). Within each chain, the atoms are strongly bonded, while the chain-chain bonding is of a weak van der Waals type. There are three unique Te sites (Te-1, Te-2, Te-3). Each Hf atom is bonded to two singly-coordinated Te-1 atoms perpendicular to the chains, while Te-2 and Te-3 are doubly coordinated to Hf and participate in the zig-zag chain structures.  We define three relevant Hf-Hf distances:  $A$ (intra-chain separation), and $B, C$ (inter-chain separations roughly along conventional unit cell axes $b, c$, respectively, see Fig.\ \ref{fig:theory-defects}(a)).

The results of our full structural optimizations with vacancies are summarized in Fig.\ref{fig:supFIGSTRcal}. We find that Te-2 is the lowest energy defect site; in comparison, Te-1 and Te-3 have energies 2.29 meV and 8.54 meV higher per atom, respectively. Te-1 vacancies reduce the $A$ intra-chain separation and $C$ inter-chain separation by about 3\%, while the Te-2 vacancy has minimal effect on those two distances. Both vacancies lead to a slight increase in the Hf-Hf $B$ inter-chain separation. Since both vacancies are overall quite close in energy, and much more favorable than the Te-3 vacancy, we can conclude that defects induce a compression along $a$, accompanied by a slight increase in the Hf-Hf $B$ inter-chain separation and another reduction in the Hf-Hf $C$ inter-chain separation. 

\begin{table}
\caption{\label{tab:z2} The $\mathbb{Z}_2 = (\nu_0;\nu_1\nu_2\nu_3)$ topological invariants and classification of gaps (I) and (II) at three different levels of strain. $\nu_0$ is the strong invariant, and $\nu_{i=1,2,3}$ are the three weak invariants \cite{Fu2007a,Fu2007b}. The weak invariants quoted here are with respect to the primitive reciprocal lattice vectors defined in Fig. \ref{fig:theory-defects}(b).}
\centering
\begin{tabular}{l*{6}{>{\centering\arraybackslash}p{4em}}}
\toprule
    {} & \multicolumn{2}{c}{2\% Compressive} & \multicolumn{2}{c}{Equilibrium} & \multicolumn{2}{c}{2\% Tensile} \\
    \cmidrule[0.6pt](lr){2-3}\cmidrule[0.6pt](lr){4-5}\cmidrule[0.6pt](lr){6-7}
            Gap & $\mathbb{Z}_2$ & Class & $\mathbb{Z}_2$ & Class & $\mathbb{Z}_2$ & Class\\
    \cmidrule[0.6pt](lr){2-3}\cmidrule[0.6pt](lr){4-5}\cmidrule[0.6pt](lr){6-7}
        I & (1;110) & Strong & (1;110) & Strong & (0;110) & Weak\\
        II & (0;110) & Weak & (0;110) & Weak & (0;110) & Weak\\
        \bottomrule
\end{tabular}
\end{table}

Considering the association of Te vacancies with compaction of the Hf-Hf inter-chain distance $A$, we hypothesize that compressive uniaxial strain along the crystallographic $a$ direction is a proxy for Te vacancies, or conversely, tensile uniaxial stress along $a$ may suppress the effects of Te vacancies. To simulate strain along the $a$-axis in our calculations (i.e., along the Hf-Hf chain direction), we artificially fixed the $a$ lattice parameter, and found the corresponding optimal lattice parameters for $b$ and $c$ for each value of strain considered (2\% uniaxial compressive and tensile strain). For all cases, full internal coordinate optimizations were performed. 

The bulk primitive electronic structures at the three different levels of strain are shown in Fig. \ref{fig:theory-defects}(d-f). We find two topological gaps of interest, labeled (I) and (II), whose $\mathbb{Z}_2$ topological invariants are summarized in Table \ref{tab:z2}. Gap (I) is a strong gap in the -2\% (compressive) and 0\% (equilibrium) strain cases, and a weak gap in the case of +2\% (tensile) strain. Gap (II) is a weak gap at all strain levels. We conclude that in the absence of defects, \hf\ undergoes a STI $\rightarrow$ WTI topological phase transition between 0 and 2\% tensile strain. Note that the 2\% compressive strain case cannot be rigorously labeled a strong topological \textit{insulator} since the bulk system is metallic, as shown in Fig. \ref{fig:theory-defects}(d). However, the band structure in Fig. \ref{fig:theory-defects}(d) can be adiabatically deformed so that no bands cross the Fermi level. Thus, \hf,\ under 2\% compressive strain is equivalent to a strong topological insulator from the perspective of band topology. We expect that \hf\ remains an insulator when smaller compressive strains are applied.
\figthree-band

A topological surface state at the $\bar{\Gamma}$ point at the Fermi energy is shown in Fig.\ \ref{fig:strain-phase} (b), (d), and (e). This state lies in gap (I), which is a strong topological gap at equilibrium and with compressive strain, so we expect an odd number of topological surface states \cite{Fu2007a,Fu2007b} (namely, one topological state at equilibrium in this system). 
With tensile strain, gap (I) becomes weak and the topological surface state at $\bar{\Gamma}$ is gapped out (Fig.\ \ref{fig:strain-phase} (c) and (f)). No topological surface states are present on the $(1\bar{1}0)$ surface in gap (I) with tensile strain, however. If we identify the surface with the $\bm{G} = (1,-1,0)$ reciprocal lattice vector, then we have $\bm{G} \mod 2 = G_\nu$, where $G_\nu = (\nu_1,\nu_2,\nu_3) = (1,1,0)$. Thus, in the weak TI state, the $(1\bar{1}0)$ surface cannot host any topological surface states \cite{Fu2007a,Fu2007b} in gap (I).

To experimentally verify the DFT predictions, we performed strain-dependent ARPES experiments on samples with different Te vacancy densities. Fig.\,\ref{fig:strain-phase} (a) shows a band structure cut along the $k_y=0$, which corresponds to the $\bar{\Gamma} - \bar{X}$ direction. The topological surface state at the Fermi energy is not clearly resolved from the bulk states, similar to previous ARPES studies for \zr\ and \hf\ \cite{Wu2016prx,Luca2016,liu2018}, but overall band structures are amazingly well matched with the DFT calculations. More detailed measurements were performed along the $\bar{\Gamma} - \bar{Y}$ direction. Figs.\,\ref{fig:band}(b-f) are band structure cuts along the high symmetry line $k_x=0$, which corresponds to the $c$-axis or $k_y$ direction perpendicular to the Hf chains. Each band structure map corresponds to a different defect density and strain, as summarized schematically in \bandref{a} for sample 1 (more defects, blue symbols) and sample 2 (fewer defects, red symbols).
\figfour-strain-phase
We first discuss the effect of defect density for unstrained samples (\bandrefs{b,e}). Although there is little observed difference between the band structures near the Fermi energy, clear differences can be observed near binding energy \eb$\sim$1 eV. Interestingly, two Dirac-like band crossings labelled $\chi_i$ ($i=1,2$), appear at $k_{y}$=$\pm$0.2$\textrm{\AA}^{-1}$ and \eb$\sim$0.95eV and 1.2 eV in \bandref{b}. These linear crossings (labelled in Fig.\ 4(d) where they also appear) are connected, respectively, by relatively flat bands $\alpha$ and a hole band $\beta$. It should be noted that, while in a repeated zone scheme $\alpha$ and $\beta$ bands are present in every BZ, matrix element effects modulate the strength of these bands such that $\alpha$ is suppressed in the central zone and $\beta$ is suppressed in the second (outer) BZs.  

A primary finding of this work is that the features \abandchi\ in \bandref{b} vanish when defect density is reduced, \bandref{e}. These results demonstrate that these features are strongly enhanced by defects in unstrained samples, a counter-intuitive result since one usually expects defects to weaken, not enhance, spectral weight. Close inspection of \bandref{e} reveals some weak remnant intensity of the $\alpha$, $\beta$ bands, but the crossings $\chi_2$ are replaced by some states around -1.3 eV with clearly different dispersion.

Now we discuss the application of uniaxial stress along the $a$ direction, i.e., in the Hf-Hf chain direction. The resulting strains are estimated to be in the range $\sim\pm$0.7\% based on strain gauge measurements with the same experimental configuration. (see Fig.\ \ref{fig:supFigStrain} and Fig.\ \ref{fig:supFigStrainBS}(a).) For a high defect density combined with tensile strain (sample 1, \bandref{c}), the \abandchi\ bands are greatly diminished, similar to unstrained sample 2 (\bandref{e}). In fact the spectral functions in \bandrefs{c,e} are practically indistinguishable, the main difference being that the spectra from the sample with more defects (1) has a higher diffuse background. On the other hand, applying compressive strain to sample 2 results in the re-appearance of the \abandchi\ bands (\bandref{d}), with spectral function very similar to \bandref{b}. 
On the other hand, in the case of tensile strain, the surface bands are completely extinguished and there is a notable rise in the diffuse background (\bandref{f}), which increases systematically towards higher binding energies. The changes in \bandrefs{d-f} are fully reversible (see Fig.\ \ref{fig:supFigStrainBS}). These changes indicate that with compressive strain, the quasiparticle scattering rate dramatically diminishes, while with tensile strain, there is a large enhancement of the quasiparticle scattering rate, suggestive of enhanced Fermi liquid-type excitations.  

From these observations we can conclude the following: (1) The similarity of \bandrefs{b,d} and of \bandrefs{c,e} confirms our hypothesis that defects play a similar role as compressive strain along the chain with respect to promoting the appearance of the states \abandchi;  (2) these bands appear in \hfd\ as a consequence of the lattice changes induced by defects and are not derived from local electronic states directly associated with the defects;  (3) defects and tensile strain independently increase the scattering rate, while compression reduces it;  (4) up to the maximum applied strain $\pm 0.7\%$ we do not increase the number of defects; this follows from the reversibility of the changes in \bandrefs{d-f} and \ref{fig:supFigStrainBS}.

 The phase diagram in \bandref{a} summarizes the observed phases. The upper left region (more defects and/or compressive strain) has a Dirac-like spectrum characterized by the appearance of the $\alpha$ and $\beta$ bands connecting linear crossings $\chi_i$ around \eb\ $\sim 1.1$ eV. The lower right region (fewer defects and/or tensile strain) is characterized by the absence of these states. Furthermore, there is a strong correlation between tensile strain applied along the $a$-axis and the quasiparticle scattering rate. The latter represents a transition towards a liquid-like quasiparticle (QP) phase, characterized by spectral weight transfer from the bands to a strong diffuse background that increases systematically with binding energy.  

To understand these features around \eb\ $\sim 1.1$ eV, we next compare our ARPES results to the DFT results. We note that the strain considered in the DFT calculations is about three times the strain on the sample in the experimental measurements, thus the changes are more pronounced between the compressive strain, equilibrium, and tensile strain band structures.  However, the equilibrium DFT calculation results (\bandref{h}) reproduce the essential features of unstrained sample 1 (\bandref{b}) and compressed sample 2 (\bandref{d}), namely the shapes and relative locations of bands $\alpha, \beta$ and crossings $\chi_{1,2}$. Since the calculation does not include photoemission matrix element effects, the bands appear equally in all BZs.

From these surface electronic structure calculations, we next interrogate the surface projected bands for the presence of topologically protected states. Although both $\chi_{1,2}$ are suggestive of Dirac crossings, neither of them are predicted by DFT to be a topological surface state. Since gap (II) is a weak gap, we expect an even number of topological surface states \cite{Fu2007a,Fu2007b}. However, since we have $\bm{G}_\nu = (\nu_1,\nu_2,\nu_3) = (1,1,0)$, similar to gap (I), no topologically non-trivial surface states can be expected on the $(1\bar{1}0)$ surface of \hf\, in gap (I) for the same reason previously discussed.

Considering the $\alpha, \beta$ and crossings $\chi_{1,2}$ bands are not directly related to the TI surface states, the enhancement of the spectral weight of the bands with more defects is counter-intuitive. Thus, it leads us to connect with the topological phase transition near the Fermi energy. For TI surface states at/near the Fermi energy, the interplay between topology and disorder effects has been well studied\,\cite{Fu2007b}: surface states of a strong/weak TI are robust/vulnerable against charge disorders and thus the surface is expected to be conductive/insulating. In contrast, at energy far away from the Fermi energy (e.g., in the gap between two valence bands), impurities also have some nontrivial impact on surface states, through a different mechanism: broadening of line width due to disorder potentials. In contrast to a homogeneous electric potential, which shifts the energy of all bands by a constant amount, random potentials created by charge impurities provide random shifts, which results in a broadening of the line width. For the surface of a strong TI, the conducting topological surface state screens the Coulomb potential from charge impurities, reducing the disorder potential and thus leading to a sharper spectrum. This effect applies equally to surface states both near and away from the Fermi energy. 
In contrast, for weak TIs, such screening effects are absent, and thus the spectrum becomes more diffusive.  As a result, if a material undergoes a topological phase transition from a strong TI to a weak TI, this effect implies that surface states away from the Fermi energy shall suffer more from impurity potential in the weak TI phase. This surface screening effect will logically affect the surface states most strongly. This is largely consistent with what was observed in the ARPES measurement, where the surface states ($\alpha, \beta$ and crossings $\chi_{1,2}$) become diffusive in the weak TI phase, while bulk states at the same energy range remain largely unchanged.

What remains to be explained is the reversible correlation between the topological class and the notable changes to the diffuse scattering background, which reflect coherence-limiting scattering processes that occur during photoemission.  While a quantitative explanation for this effect is beyond the scope of this study, we can suggest three qualitative explanations: (1) the screening near the surface is reduced by the absence of metallic topological states in the WTI phase, thus the scattering of bulk states near the surface (that dominate the ARPES signal) is enhanced in the WTI phase. This leads to the QP-like increase in scattering rates towards higher binding energy in the WTI phase; (2) tensile strain reduces the electron density along the Hf chains, thus similarly reducing the bulk screening and enhancing the QP scattering rate in the bulk; (3) defects such as dislocations are created upon tensile strain and reversibly destroyed when strain is relieved or compressive strain applied.  (We find this explanation unlikely given that we are below the plastic deformation regime, but we mention the possibility for completeness.)

Since we can assume that the diffuse background is dominated by the QP scattering of bulk state\cite{cai_independence_2018}, the key to distinguishing effects (1) and (2) is the behavior of the momentum broadening of the bulk states with strain. For weakly-correlated systems, the bulk states' momentum width \dk\ at binding energy $\omega$ from ARPES is directly proportional to the imaginary part of the self-energy function, which reflects the electron lifetime to all scattering from phonons, defects, or other electrons.  For (1), we would expect an abrupt general increase in \dk\ as we transition from STI to WTI, and for (2) we would expect a continual decrease in  scattering as we go from tensile to compressive strain, regardless of topological phase transition.

While our analysis of the  bulk states' momentum spread \dk\ has not revealed a clear evolution with strain (see Fig.\ \ref{fig:supFigArpesFit}), favoring explanation (1), it remains possible that the changes in \dk\ are below our experimental resolution, or that the observed \dk\ is dominated by inhomogeneous broadening effects.

Nevertheless, our results strongly suggest that the self-energy of both surface and bulk electronic (trivial) states are strongly impacted by the presence of topological surface states, an effect not seen before in ARPES to our knowledge. Our results further demonstrate strain control of topological states as a new method to tune or optimize the coherence of electronic excitations, even overcoming scattering by unavoidably present defects, to improve the performance of quantum materials in new device schemes.

\newpage

\section{Data availability}
Data underlying these results are available from the authors, see Additional Information below. 

\section{References and notes}

\normalem
%

\section{Acknowledgement}

This material is based upon work at the QSA, supported by the U.S. Department of Energy, Office of Science, National Quantum Information Science Research Centers. This research used resources of the Advanced Light Source, which is a DOE Office of Science User Facility under contract no. DE-AC02-05CH11231. This work was also supported by the DOE's Quantum Information Science Enabled Discovery (QuantISED) for High Energy Physics (KA2401032). Computational resources were provided by the National Energy Research Scientific Computing Center and the Molecular Foundry, DOE Office of Science User Facilities supported under Contract No. DEAC02-05-CH11231. The work performed at the Molecular Foundry was supported by the Office of Science, Office of Basic Energy Sciences, of the U.S. Department of Energy under the same contract. O.A.A. acknowledges helpful discussions with I. Na and E. Banyas.

\section{Author contributions}
N.H.J. and E.R. conceived of the presented idea. A.S. and T.K. synthesized the single crystal and performed X-ray diffraction and electrical transport measurements. N.H.J, C.J., A.B., S.H.R, and E.R. conducted the ARPES experiment. O.A.A. and S.M.G. performed the DFT calculations and topological characterization. O.A.A., K.S., and S.M.G. carried out theoretical analysis. N.H.J., O.A.A., K.S., T.K., S.M.G., and E.R. wrote the manuscript in consultation with Z.S., C.J. A.B., and S.H.R.

\section{Competing financial interests}
The authors declare no competing financial interests.

\section{Additional information}
Correspondence and requests for materials should be addressed to N. H. Jo, S. M. Griffin, and E. Rotenberg. 

\section{Methods}
\textbf{Crystal growth.} Single crystals of \hf\ were grown using tellurium flux. Hafnium powder (Alfa Aesar, 99.6 $\%$ metals basis excluding Zr, Zr nominal 2-3.5 $\%$) and a tellurium lump (Alfa Aesar, 99.999$\%$), in the atomic ratio of Hf:Te = 1:99, were sealed in a silica tube under vacuum. The growth was heated up to 650 $^{o}$C in 3 hours, dwell for 10 hours, and then slowly cooled to 460$^\circ$C over 78 hours before decanting. Low temperature annealing was conducted at 380$^\circ$C and 250$^\circ$C for 5 days in a sealed silica tube for sample 1 and sample 2, respectively. The samples crystallize in space group number 63 ($Cmcm$). The lattice parameters for the single crystals are $a~\sim$ 3.97~\AA${}^{-1}$, $b~\sim$ 14.5~\AA${}^{-1}$, and $c~\sim$ 13.7~\AA${}^{-1}$. 

\textbf{Electrical transport measurements.} Resistance measurements down to 1.8 K were conducted using a Quantum Design physical property measurement system (PPMS) Dynacool using the resistivity option. Pt wires were attached to the samples using DuPont silver paint (4929N) in a standard 4-probe configuration. Magnetic field (0-90 kOe) was applied perpendicular to $a$-axis. 

\textbf{ARPES measurements.} ARPES experiments were conducted at the Beamline 7.0.2 (MAESTRO) at the Advanced Light Source. The data were acquired using the micro-ARPES endstation, which consists of an Omicron Scienta R4000 electron analyzer. Samples were cleaved $in~situ$ by carefully knocking off an alumina post that is attached on top of each sample with silver epoxy. Data were collected with photon energies of 101~eV, which accesses near the $\Gamma$ plane of the Brillouin Zone (BZ). The beam size was $\sim$~15$~\mu$m~$\times$~15$~\mu$m. ARPES measurements were performed at $T=$200~K under ultra-high vacuum (UHV) better than 4~$\times$~10$^{-11}$ torr. 

As shown in Fig.\ \ref{fig:theory-defects}(a), \hf\ has a weak van der Waals bonding along the crystallographic $b$ direction; thus, the cleaved surface was $ac$-plane. Note that, by ARPES convention, $k_{z}$ is normal to the cleaved surface. Therefore, we define $x$, $y$, and $z$ as directed along the crystallographic $a$, $c$, and $b$, respectively.

\textbf{Uniaxial stress.} ARPES measurements under uniaxial stress were carried out using a custom-built uniaxial stress cell, as shown in Fig.\ \ref{fig:supFigCell}. Samples were mounted with Stycast epoxy (2850FT). After curing the epoxy, we put silver paste (Dupont, 4929N) on top to ground the sample. The uniaxial stress was applied on samples along the crystallographic $a$ direction. In this experimental configuration, three strain tensor components were nonzero in Voigt notation\cite{Jo2019}. We measured the strain tensor component $\epsilon_{xx}$ with a strain gauge (Fig.\ \ref{fig:supFigStrain}), and the maximum tensile/compress strain was $\sim\pm$0.7\%. Detailed information about the cell can be found in SI. In order to keep the same beam spot on the sample, we used a microscope in the $\mu$-ARPES chamber. In addition, we performed $xy$ scan on the sample with 10$\mu$m step size and slit-deflector scan each and every time we change the strain on the sample. Note that we obtained the unstrained data after cooling down the cell but before applying any voltages on the piezo actuators. Even though no voltages are applied, the thermal contraction of the cell, which is most likely different from the thermal contraction of the sample, does slightly exist. 

\textbf{DFT calculations.} Our DFT calculations were performed using the Vienna \textit{Ab initio} Simulation Package (\textsc{vasp}) \cite{Kresse1993,Kresse1994,Kresse1996,Kresse1996b} using the projector augmented wave (PAW) method \cite{Kresse1999}. All calculations in the main text use the primitive cell, except those for Fig.\ \ref{fig:supFIGSTRcal} which used a $2\times2\times2$ supercell of 96 atoms. Hf(5p, 6s, 5d) and Te(5s, 5p); electrons were treated as valence. We expanded the wavefunction plane waves to an energy cutoff of 600 eV, and used $\Gamma$-centered k-point grids of $10\times 8 \times 6$ for the conventional cell, $12\times 12\times 6$ for the primitive cell for structural optimizations, and $4\times 8\times 2$ for the supercell defect structural optimizations. An $18\times14 \times 10$ grid was used with the conventional cell for accurate electronic structure calculations and to generate the Wannier-based tight-binding model. We did not include spin-orbit coupling in our structural optimizations as it was found to have minimal influence on the calculated lattice parameters, however, we included spin-orbit coupling self-consistently in the electronic structure calculations where specified. The electronic convergence criterion is set to $10^{-7}$ eV and the force convergence criterion is set to 0.005 eV / \AA{}. 

Prior DFT calculations on \zr\ and \hf\ highlighted the importance of accurate calculations of the structural parameters, particularly owing to the extreme sensitivity of details of the electronic and topological properties on the volume and structure \cite{Hochberg_et_al:2018}. Therefore, we carefully chose our exchange-correlation functional by benchmarking against reported low-temperature X-ray diffraction measurements, finding PBEsol to give good agreement with the experiment. We calculated the lattice parameters to be a = 3.953 \AA, b = 14.564 \AA, and c = 13.622 \AA\ using PBEsol, compared to the measured values of 3.964 \AA, b = 14.443 \AA, and c = 13.684 \AA at 10 Km\cite{FJELLVAG1986}.

For the slab calculations, an 11-layer (132-atom) centrosymmetric slab with 15 \AA{} of vacuum was utilized. We used a $\Gamma$-centered $6\times 20 \times 1$ k-grid, and spin-orbit coupling was included self-consistently. Line thickness and color scale in the slab band structures (Fig. \ref{fig:band} (b-f)) are given by Eq. \eqref{eq:weighted_projections} below
\begin{align}
   W_{n\bm{k}} =  \sum_{\tau_\perp} \sum_{\bm{\tau}_\parallel} \sum_{lm} |\bra{\beta_{lm}(\bm{\tau})}\ket{\psi_{n\bm{k}}}|^2 e^{-\alpha \tau_\perp},
   \label{eq:weighted_projections}
\end{align}
where $\ket{\psi_{n\bm{k}}}$ is the Bloch function of band $n$ at k-point $\bm{k}$. $\ket{\beta_{lm}(\bm{\tau})}$ is a localized atomic state with orbital quantum numbers ($lm$), centered at an atom with coordinates $\bm{\tau} = (\bm{\tau}_\parallel, \tau_\perp)$, where $\tau_\perp$ is the non-periodic direction of the slab. For a given band $n$ and $k$-point $\bm{k}$ in the band structure, we first sum over all quantum numbers ($lm$) of atomic states $\ket{\beta_{lm}}$, from all atoms with coordinates $\bm{\tau}_\parallel$ on a given plane $\tau_\perp$. We finally sum over all the planes with various $\tau_\perp$, with an exponential weighting factor $e^{-\alpha \tau_\perp}$ that decays away from the surface of the slab at $\tau_\perp = 0$. The last sum is performed over the top half of the centrosymmetric slab, giving the weighted projections from one of the surfaces at $(n,\bm{k})$. Here, we used $\alpha = 1/5$ \AA$^{-1}$.

Using our first-principles electronic structure, we constructed our Wannier function-based tight-binding models using the \textsc{WANNIER90} package\cite{Wannier90}, including the Hf-\textit{d} and Te-\textit{p} orbitals as basis. Besides the explicit DFT slab calculations mentioned above, we further calculated the surface states by generating a semi-infinite slab configuration from our tight-binding model using the iterative Green's function (IGF) method to generate surface Te-projected bands using the WannierTools \cite{WannierTools} package. Finally, our topological invariants were calculated using symmetry indicators and DFT-calculated eigenvalues as implemented in the \textsc{symtopo} \cite{Symtopo} package, \textsc{irrep} \cite{irrep, bilbao} package and the \textsc{Check Topological Mat.} tool on the Bilbao Crystallographic Server \cite{vasp2trace}. 

Gaps (I) and (II) were checked across the entire irreducible Brillouin zone (IBZ) at all levels of strain to ensure the bands do not touch (i.e., the gap remains open) using a $k$-grid of 36$\times$36$\times$12. In all cases but one, the minimum gap size was on the order of 10 meV. However, the bands got closer than 10 meV in the case of gap (II) at equilibrium, so it was further checked with a dense cubic `patch' centered around the k-point with the gap minimum. This calculation showed that the bands almost touch within ~1 meV, which is beyond the resolution of DFT. Thus, gap (II) practically closes at this level of strain at the DFT level. However, since gap (II) is finite for both compressive and tensile strain, we expect any small deviation from the equilibrium structure calculated in DFT to result in gap (II) opening.

\newpage

\include{main_suppfig_defns}

\include{main_supplemental}


\end{document}

%% file: main_figure_definitions.tex
\def\figone{
    \begin{figure} [h]
    	\includegraphics[width=3 in]{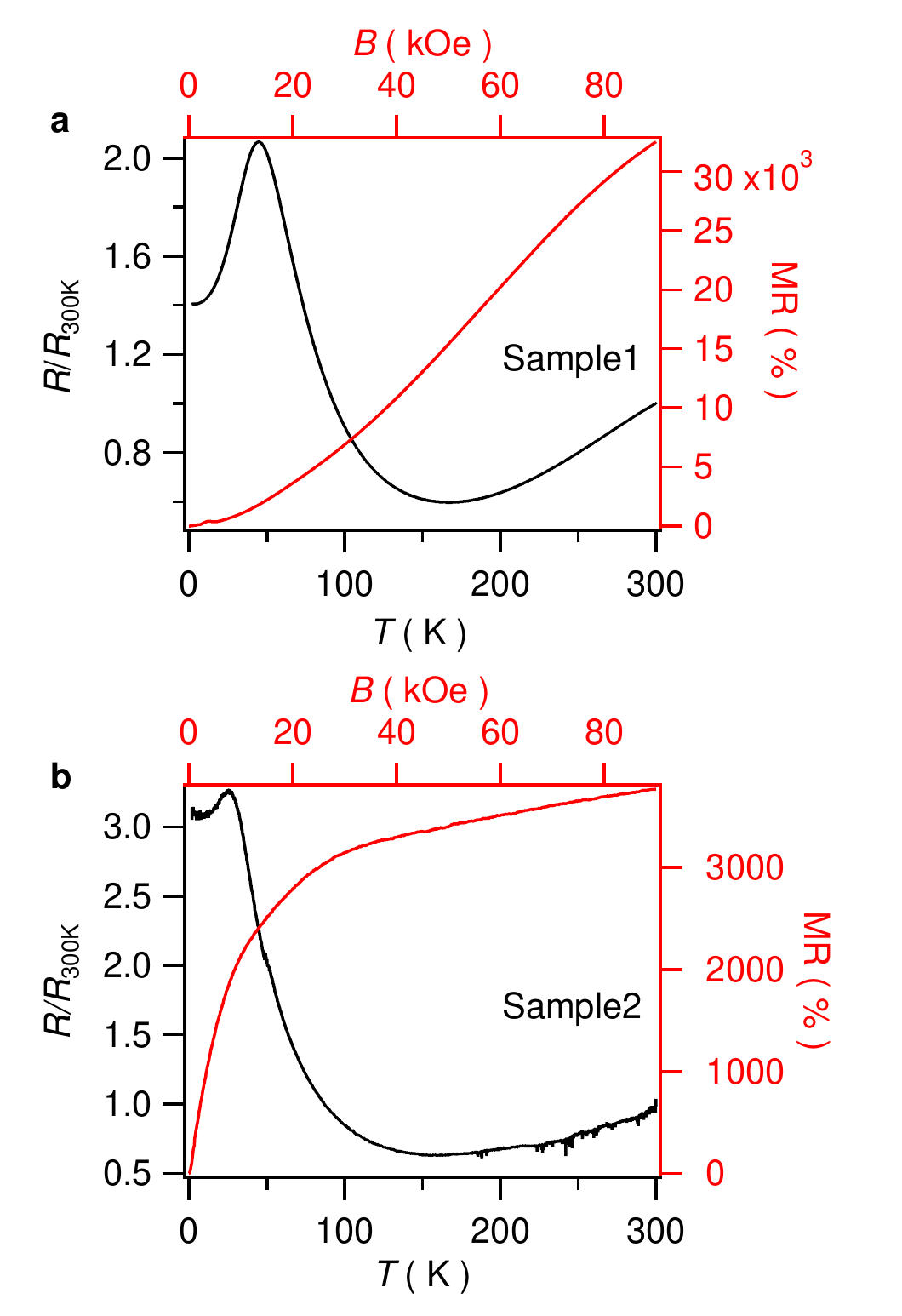}%
    	\caption{\textbf{Electrical transport.} 
    	(a) Temperature-dependent resistance ratio (left and bottom axis) and MR (right and top axis) of \hf\ sample 1 that is post-annealed at 380~$^{o}$C. 
        (b) Temperature-dependent resistance ratio (left and bottom axis) and MR (right and top axis) of \hf\ sample 2 that is post-annealed at 250~$^{o}$C.
    	\label{fig:Fig1_1}}
    \end{figure}
}

\def\figtwo-theory-defects{
    \begin{figure*} [h]
    \includegraphics[width=6in]{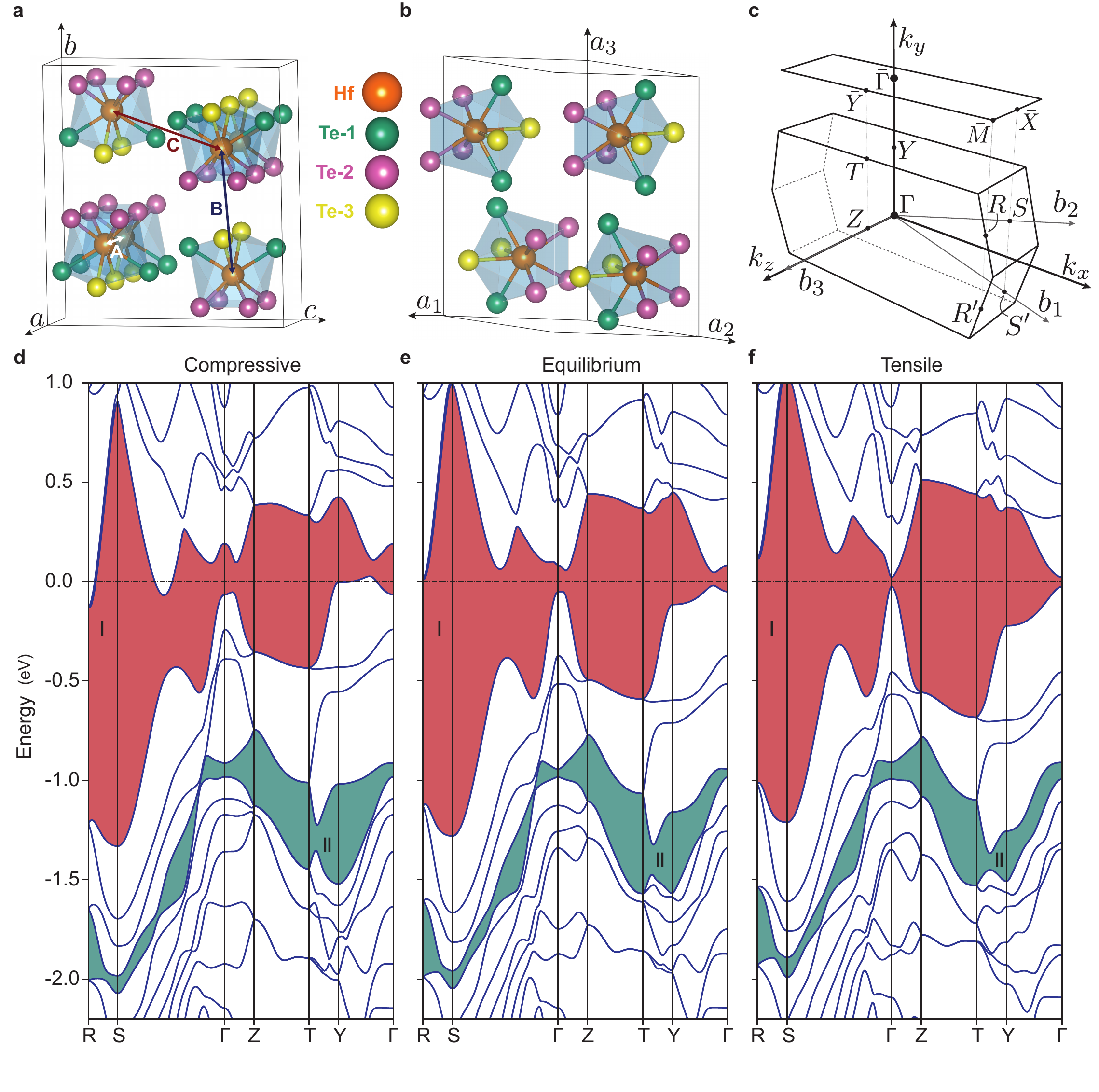}
        \caption{\textbf{Effect of Te vacancies and strain on the crystal structure and electronic structure of \hf} 
    	(a) The conventional cell of \hf, depicting the three unique Te sites and Hf-Hf bonds. (b) The primitive cell of \hf. The primitive cell has half the volume and number of atoms of the conventional cell, and was used for all bulk DFT calculations. 
    	(c) The Brillouin zone (BZ) of the primitive cell, along with the BZ of the $(1\bar{1}0)$ surface. $b_i$ are the reciprocal lattice basis vectors of the bulk BZ, corresponding to $a_i$ in (b).
    	(d)-(f) Calculated electronic band structures of \hf\ under uniaxial compressive strain, at equilibrium, and under uniaxial tensile strain, as defined in the main text. The two topological gaps of interest are shaded and labeled as (I) and (II). The Fermi level in each is set to 0 eV and marked by a dashed line. 
        \label{fig:theory-defects}}
    \end{figure*}
}

\def\figthree-band{
    \begin{figure*} [h]
    	\includegraphics[width=6.5 in]{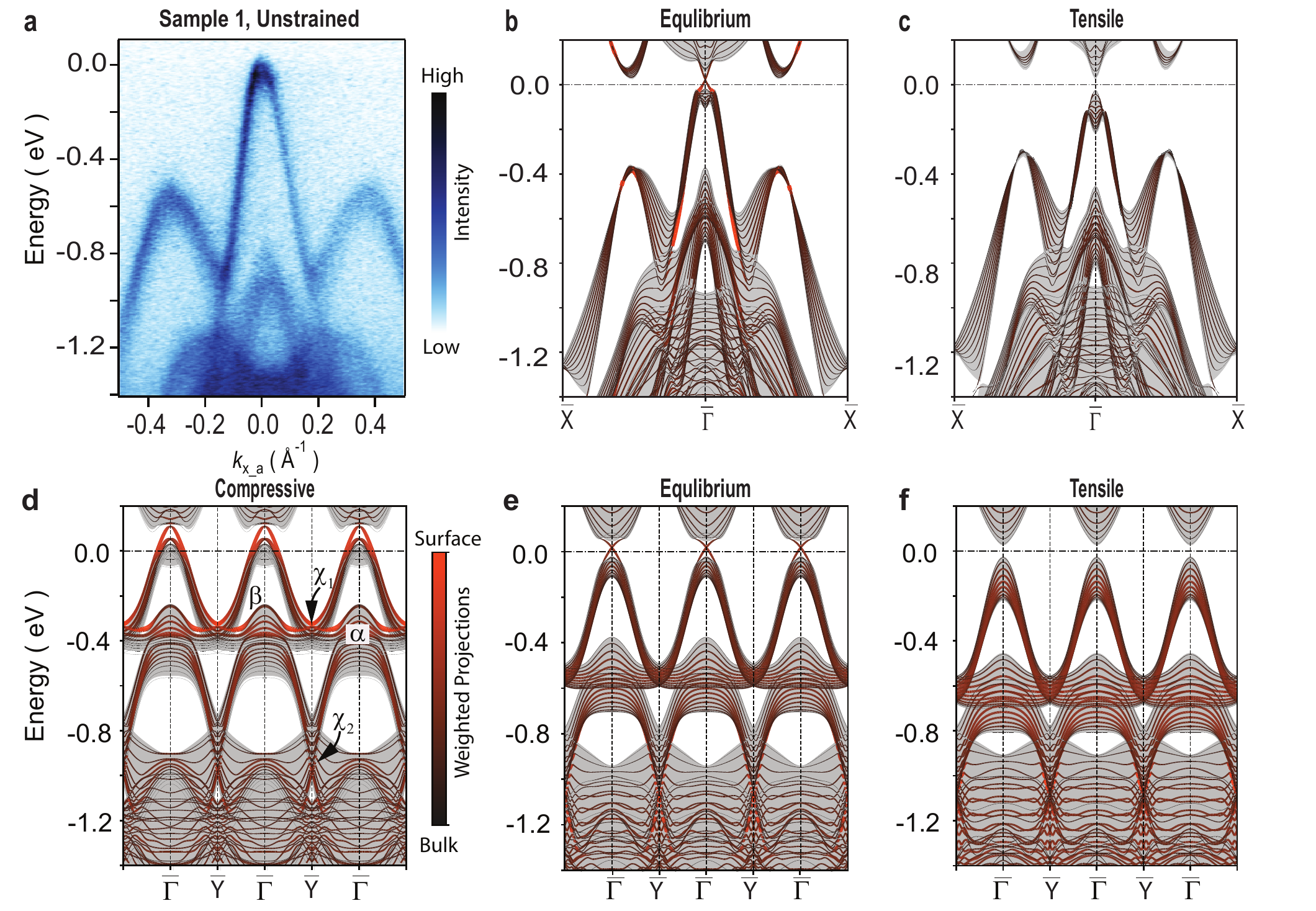} 
    	\caption{\textbf{Non-trivial topological states at $\bar{\Gamma}$}
    	(a) Electronic band structure along the $k_{x}$ of sample 1 at ambient condition. 
    	(b-c) Slab band structures along the $\bar{\Gamma} - \bar{X}$ direction for equilibrium, and tensile strain, respectively.
    	(d)-(f) Calculated surface band structure along the $\bar{\Gamma} - \bar{Y}$ direction of pristine \hf\ with compressive strain, at equilibrium, and with tensile strain, respectively. The gray shading represents bulk bands projected onto the $(1\bar{1}0)$ surface, while the colored lines represent the slab bands. The color and line width of the slab bands are given by Eq. \eqref{eq:weighted_projections} in the supplementary information, with thin black lines representing bulk character and thick red lines representing surface character. 
    	\label{fig:strain-phase}}
\end{figure*}
}

\def\figfour-strain-phase{
    \begin{figure*} [h]
    	\includegraphics[width=6 in]{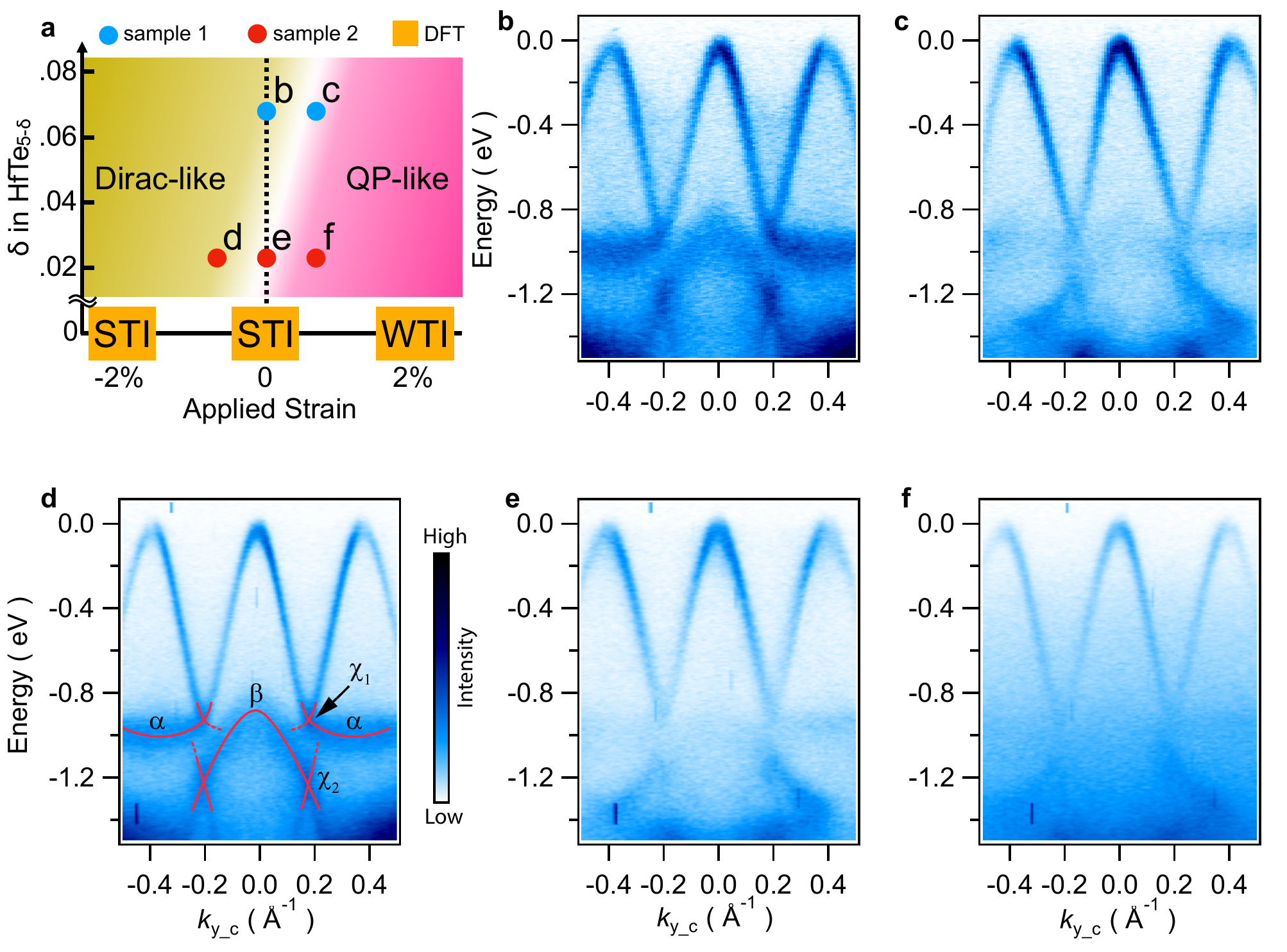}%
    	\caption{\textbf{Topological phase diagram and electronic band structure along $k_{y}$ with different Te vacancies and strain}
            (a) Schematic phase diagram, indicating the location of ARPES band structure measurements in the space of defect density \textit{vs.} strain. Blue markers correspond to Sample 1 and (b,c); Red markers correspond to Sample 2 and (d-f). 
    	(b-f) ARPES band structures acquired for positions indicated in (a).
    	\label{fig:band}}
    \end{figure*}
}

\def\bandref #1{Fig.\ \ref{fig:band}(#1)}
\def\bandrefs #1{Figs.\ \ref{fig:band}(#1)}

%% file: main_suppfig_defns.tex
\def\supfigcell{
    \begin{figure*}[h]
    	\includegraphics[width=3 in]{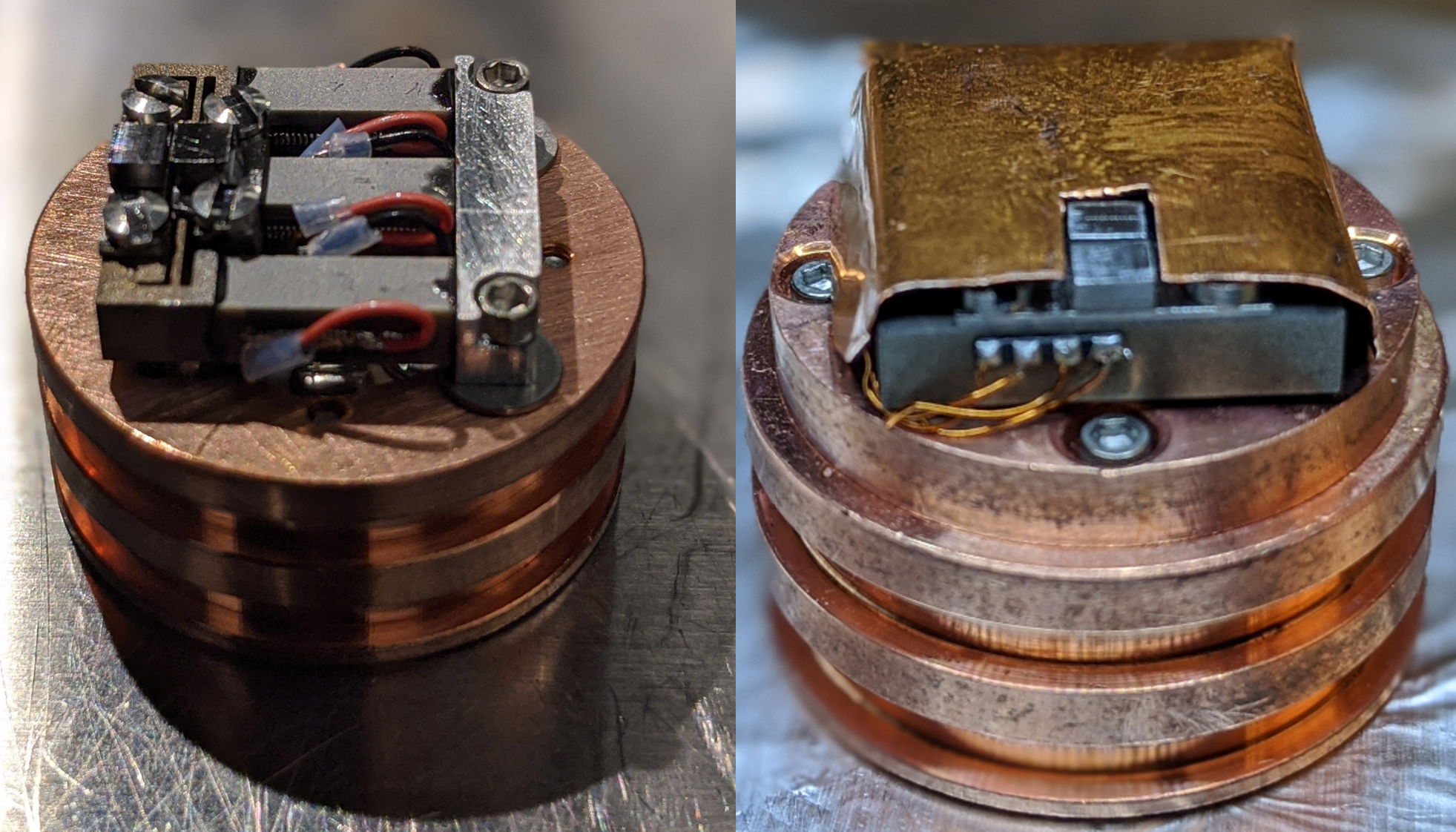}%
    	\caption{\textbf{Uniaxial stress cell}
    	Uniaxial stress cell for ARPES.
    	\label{fig:supFigCell}}
    \end{figure*}
}

\def\supFigStrain{
    \begin{figure*}[h]
    	\includegraphics[width=6 in]{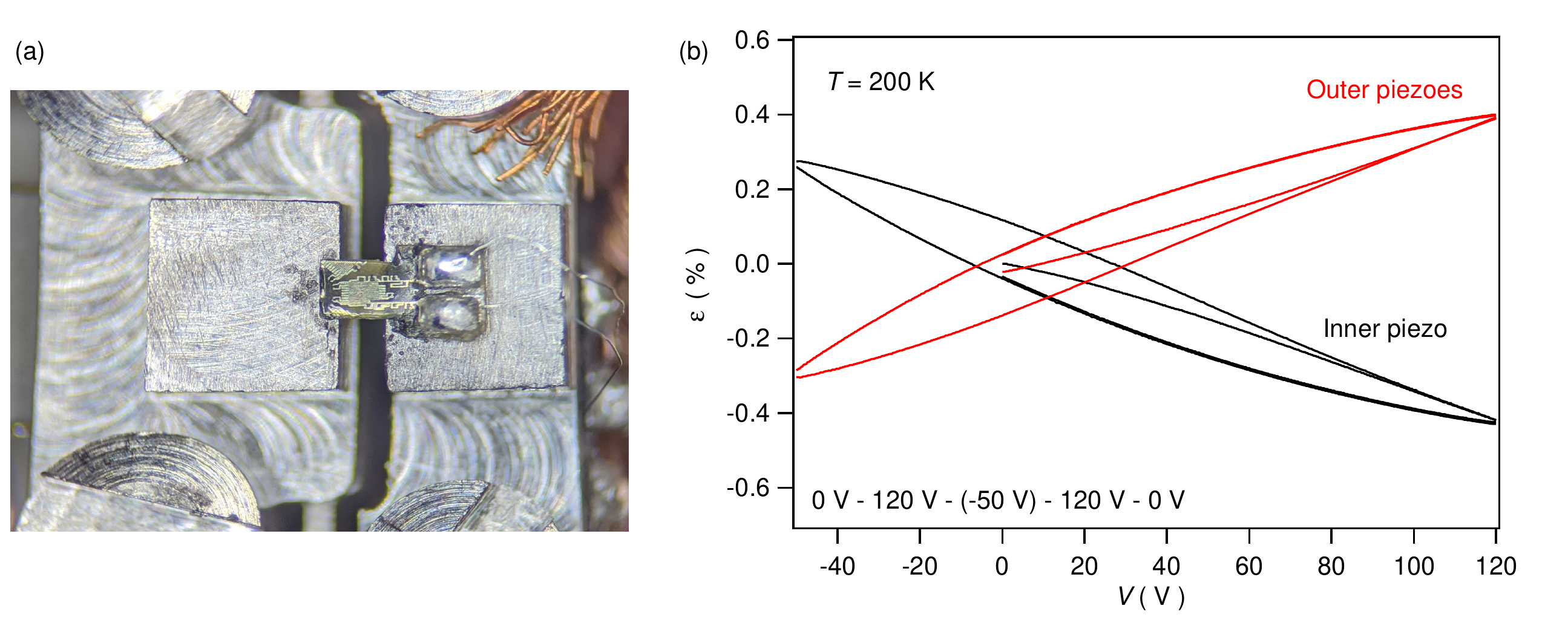}%
    	\caption{\textbf{Strain measurements via strain gauge}
    	(a) A picture of the strain gauge mounted on fixtures. (b) Strain responses as a function of applied voltage on piezo actuators.
    	\label{fig:supFigStrain}}
    \end{figure*}
}

\def\supFigResistivity{
    \begin{figure*}[h]
        \includegraphics[width=0.8\linewidth]{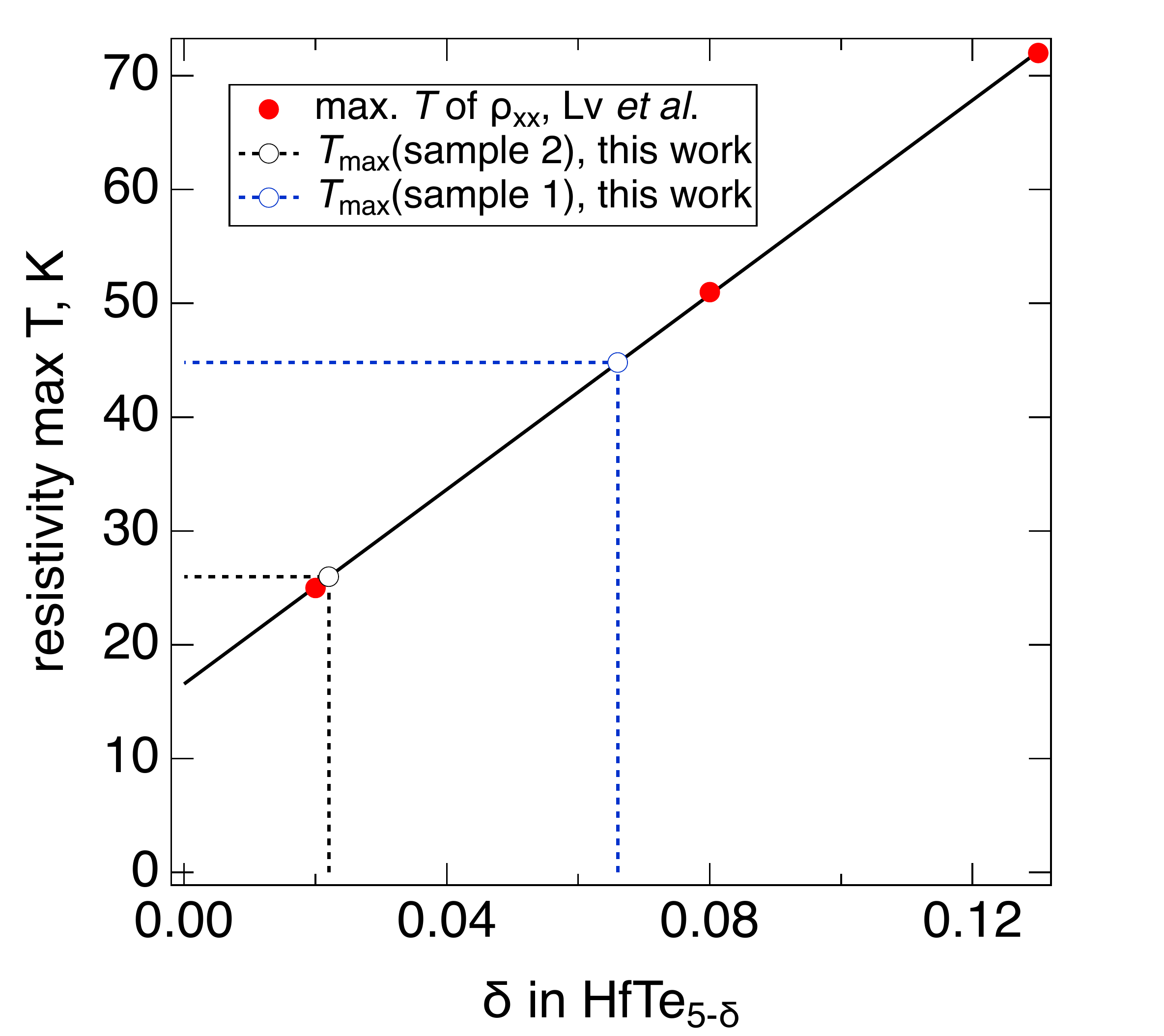}%
    	\caption{\textbf{Uniaxial stress cell}
    	Uniaxial stress cell for ARPES.
    	\label{fig:supFigResistivity}}
    \end{figure*}
    
}

\def\supFigStrainBS{
    \begin{figure*}[h]
    	\includegraphics[width=6.3 in]{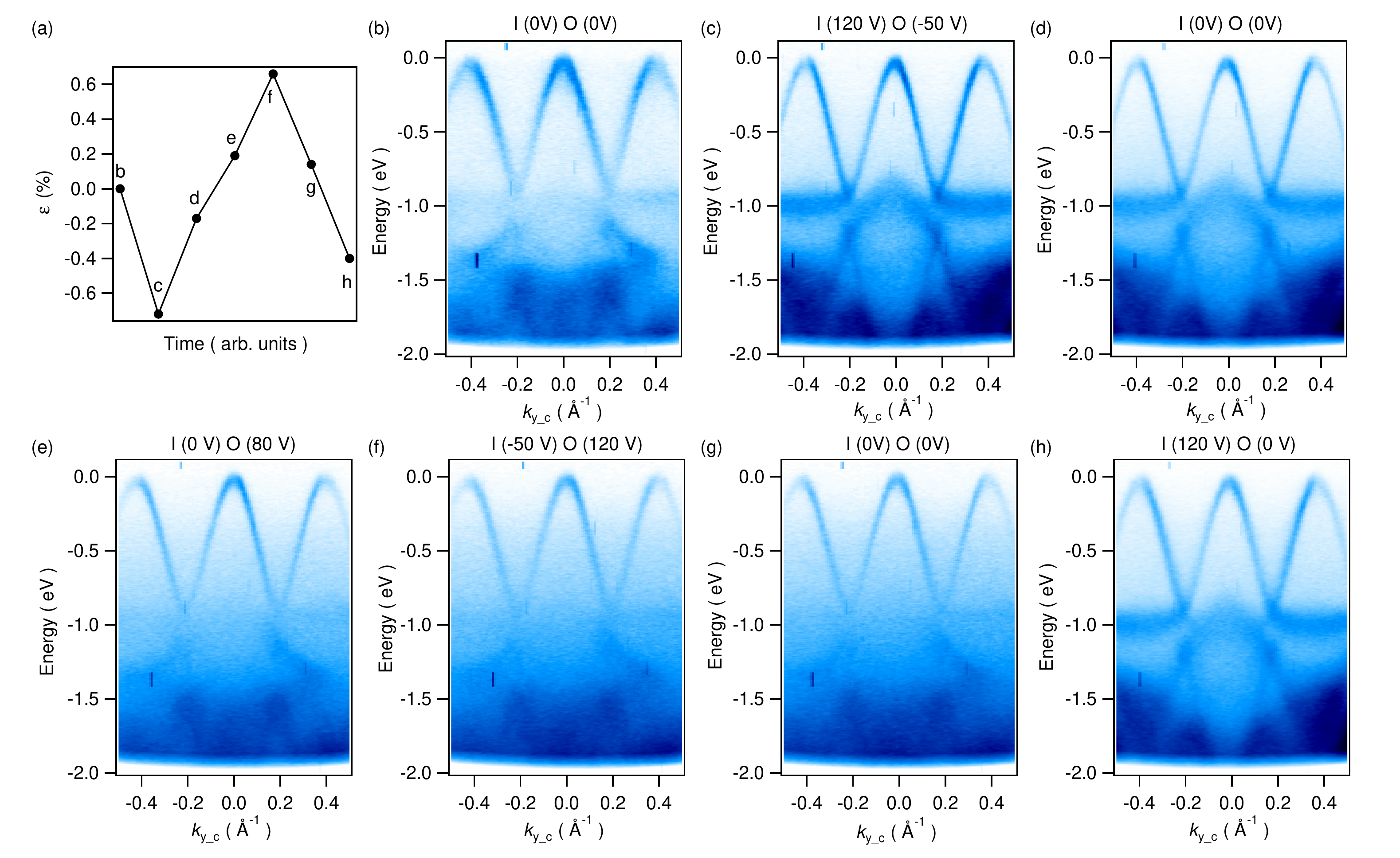}%
    	\caption{\textbf{Strain dependent band structure}
    	(a) Estimated strain based on strain gauge measurement. Band structure cut along the $\Gamma$-$Z$(crystallographic $c$) direction of \hf\ with the applied voltage: (b) inner piezo actuator of 0 V and outer piezo actuators of 0 V, (c) inner piezo actuator of 120 V and outer piezo actuators of -50 V, (d) inner piezo actuator of 0 V and outer piezo actuators of 0 V, (e) inner piezo actuator of 0 V and outer piezo actuators of 80 V, (f) inner piezo actuator of -50 V and outer piezo actuators of 120 V, (g) inner piezo actuator of 0 V and outer piezo actuators of 0 V, and (h) inner piezo actuator of 120 V and outer piezo actuators of 0 V.
    	\label{fig:supFigStrainBS}}
    \end{figure*}
}

\def\supfigisoenergy{
    \begin{figure*} [h]
    	\includegraphics[width=6.3 in]{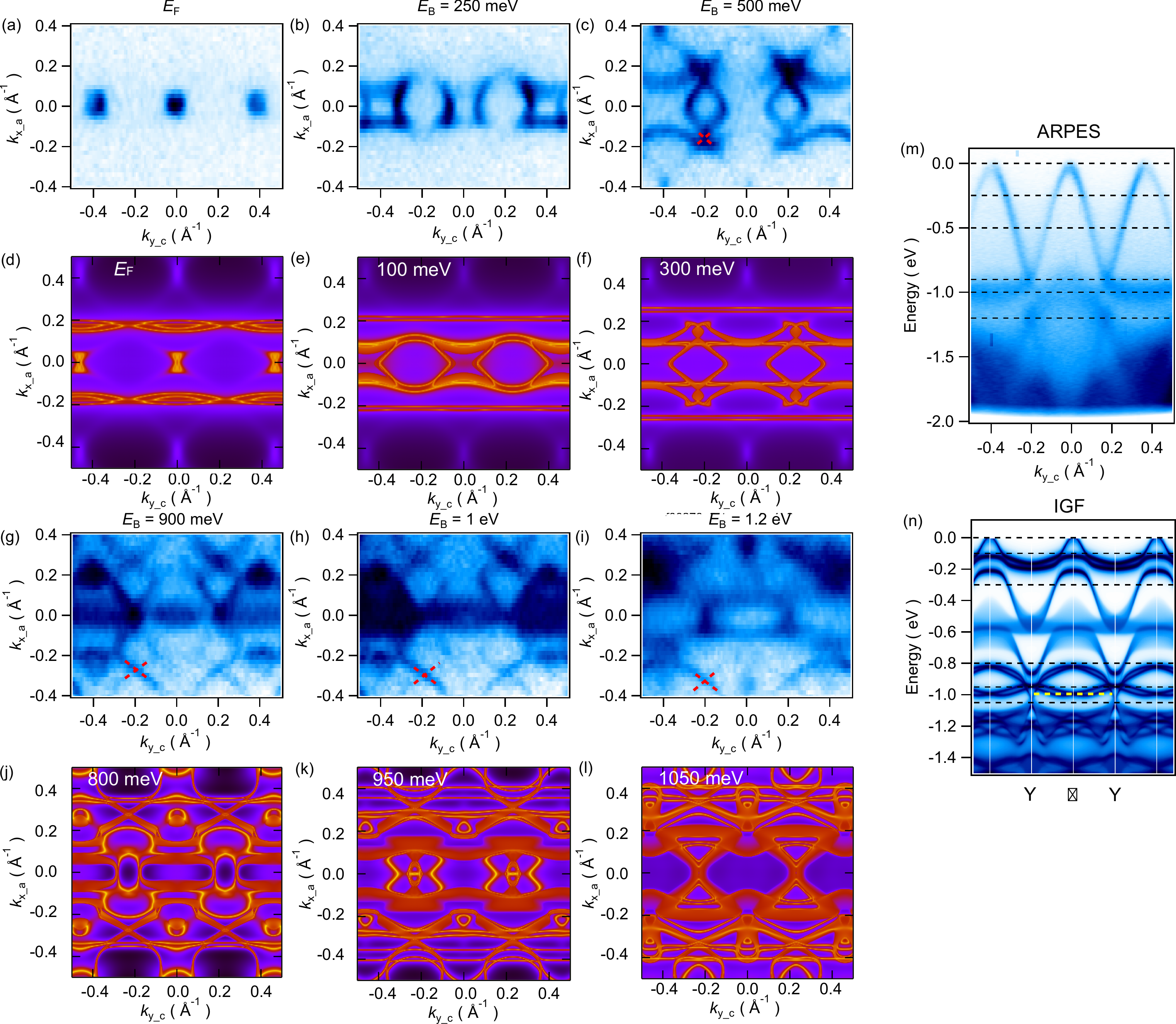}
    	\caption{\textbf{Constant energy contour plots}
    	(a)-(c) Constant energy contour plots of compressed sample 2 at $E_{F}$, $E_{B}\,=\,250$\,meV, $E_{B}\,=\,500$\,meV, respectively. (d)-(f) Constant energy contour plots from IGF calculations at $E_{F}$, $E_{B}\,=\,100$\,meV, $E_{B}\,=\,300$\,meV, respectively. (g)-(i) Constant energy contour plots of compressed sample 2 at $E_{B}\,=\,900$, $E_{B}\,=\,1$\,eV, $E_{B}\,=\,1.2$\,eV, respectively. (j)-(l) Constant energy contour plots from IGF calculations at $E_{B}\,=\,800$\,meV, $E_{B}\,=\,950$\,meV, $E_{B}\,=\,1.05$\,eV, respectively. (m) Band structure of compressed sample 2 along the $k_y$. The dashed line marks the energy from which the constant energy contour plots are taken. (n) Band structure from IGF calculations along the $k_y$. The dashed line marks the energy from which the constant energy contour plots are taken. Red dashed lines indicate another linear band crossing point.  
    	\label{fig:isoenergy}}
    \end{figure*}
}

\def\supfiglinear{
    \begin{figure*}[h]
    	\includegraphics[width=4 in]{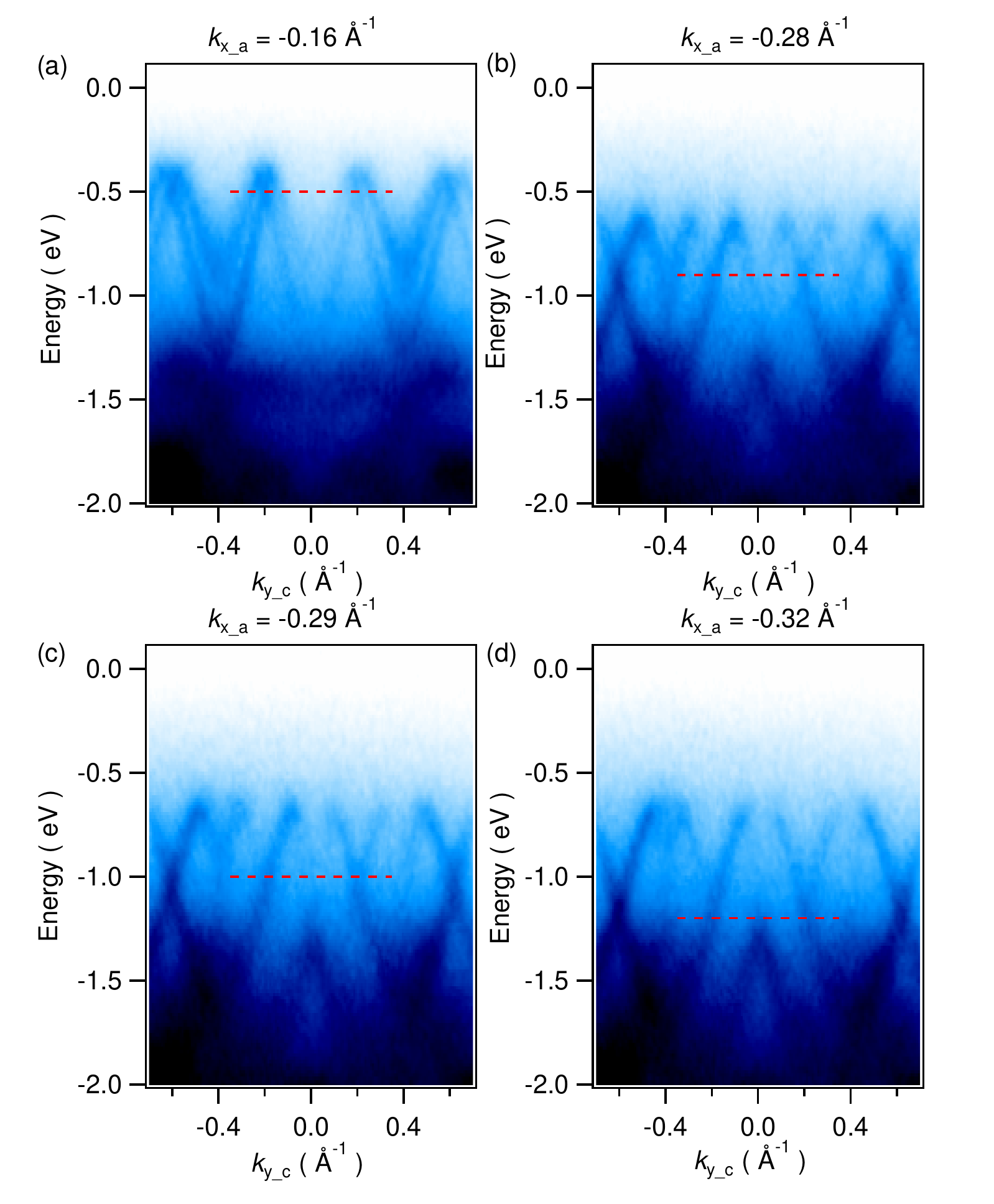}
    	\caption{\textbf{Electronic band dispersion in compressed \hf\ at the crossing points}
    	 (a)-(d) Electronic band dispersion at $k_x$ = -0.16\,\AA$^{-1}$, $k_x$ = -0.28\,\AA$^{-1}$, $k_x$ = -0.29\,\AA$^{-1}$, and $k_x$ = -0.32 \AA$^{-1}$ that are marked in Fig.\ \ref{fig:isoenergy} (c)-(f), respectively. Red dashed lines indicate the Dirac point at \eb\ = 500 meV, 900 meV, 1 eV, and 1.2 eV.
    	\label{fig:linear}}
    \end{figure*}
}

\def\supFIGSTRcal{
    \begin{figure*}[!p]
    	\includegraphics[width=0.85\linewidth]{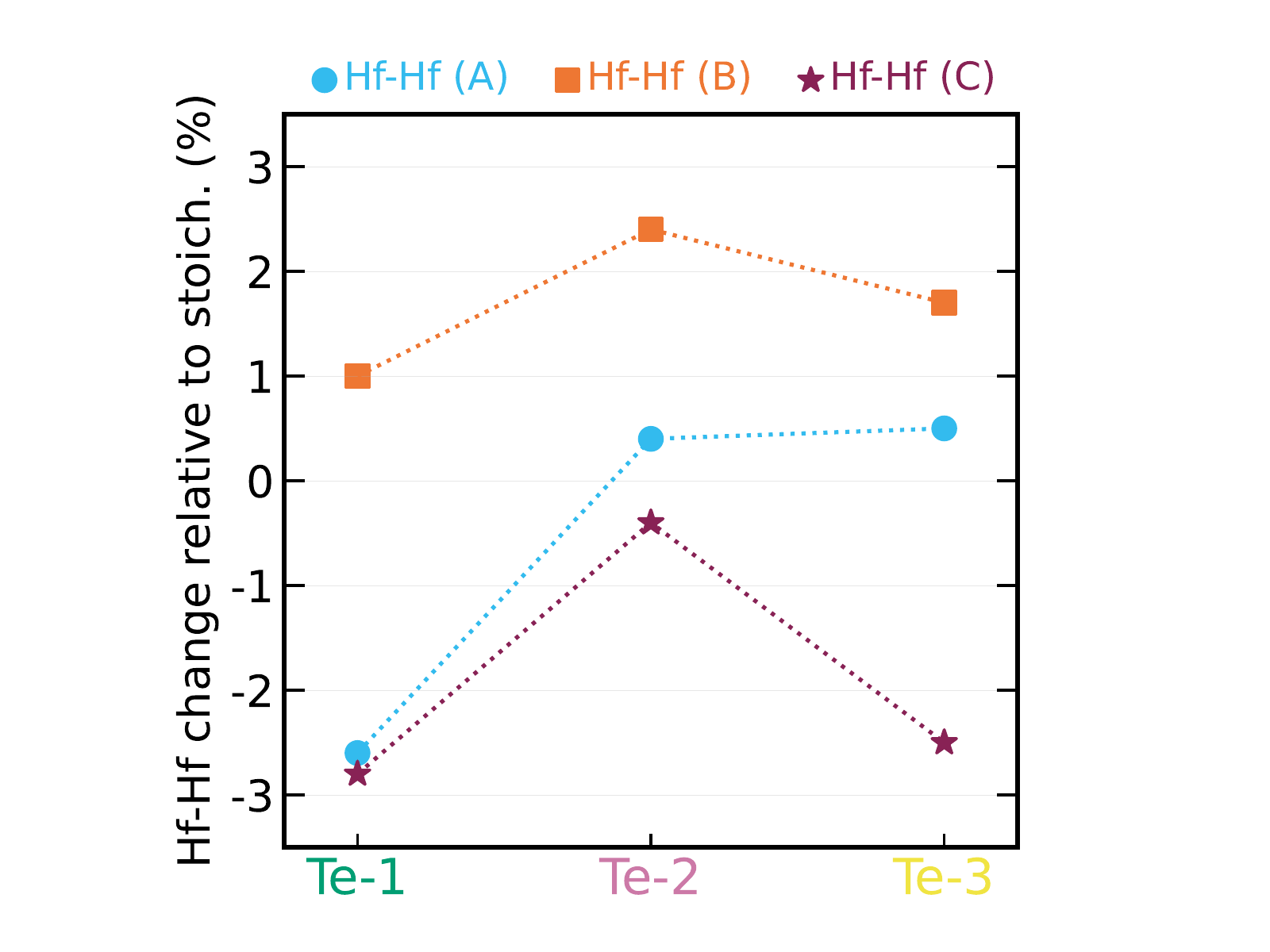}
    	\caption{\textbf{Structural changes with Te vacancies}
    	Calculated structural changes for HfTe$_{5}$ with the inclusion of Te vacancies on the three unique Te sites in the $Cmcm$ structure.
    	\label{fig:supFIGSTRcal}}
    \end{figure*}
}

\def\supFigArpesFit{
    \begin{figure*} [h]
    	\includegraphics[width=5.5 in]{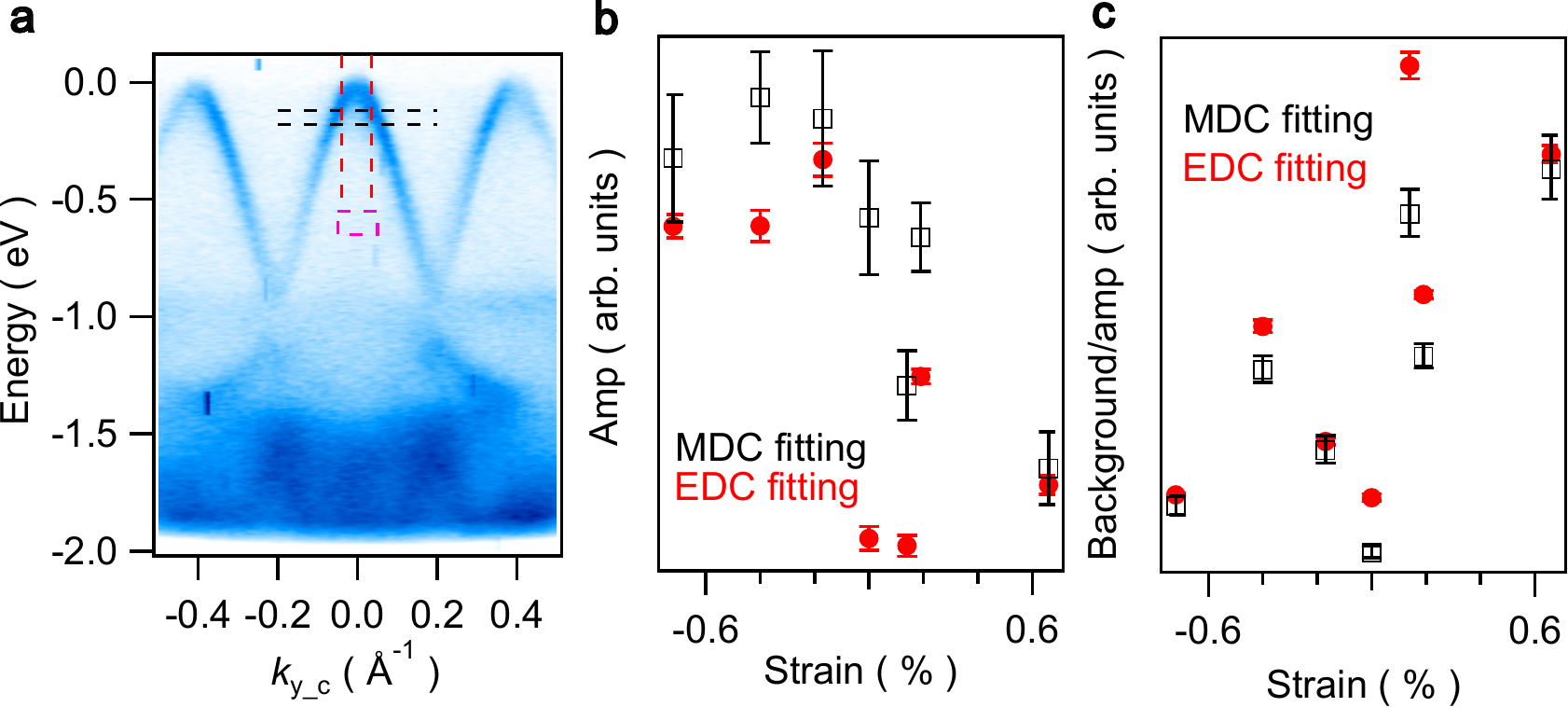}
    	\caption{\textbf{Detailed analysis of ARPES data}
    	(a) Band structure of unstrained sample 2 along the $\Gamma$ - Y direction. Black dashed lines mark the binning range for the momentum distribution curve (MDC) fitting. Red dashed lines indicate the binning range for the energy distribution curve (EDC) fitting. The magenta rectangular box shows the area where the averaged background is obtained. 
    	(b) Average amplitudes of bands based on the MDC and EDC fitting in the marked range shown in (a). 
    	(c) Normalized background signal over amplitudes of bands. The amplitudes are based on (b). \label{fig:supFigArpesFit}}
    \end{figure*}
}

%% file: main_supplemental.tex
\clearpage\pagenumbering{arabic}
\section{Supplementary Information}

\renewcommand{\thepage}{S\arabic{page}}  
\renewcommand{\thesection}{S\arabic{section}}   
\renewcommand{\thetable}{S\arabic{table}}   
\renewcommand{\thefigure}{S\arabic{figure}}
\setcounter{figure}{0}
\setcounter{table}{0}

\subsection{Uniaxial stress cell}

The uniaxial stress cell we used for this experiment is based on three piezoelectric stacks, similar to Ref. \cite{hicks2014}. As we have eight available electrical channels on our system, we used four to control the piezoelectric stacks and the rest of the four channels for strain gauge measurements. We attached a Cu thermal braiding near the sample in order to ground the sample and achieve better temperature control. In addition, we put a Cu shielding on top of the cell to shield the field from the piezoelectric stacks, as shown in Fig.\,\ref{fig:supFigCell} on the right side.
\supfigcell

\newpage\subsection{Strain measurements}
\supFigStrain
The strain was measured using a strain gauge (C5K-XX-S5198-350/39F, Micro-Measurements.) mounted with the Stycast epoxy (2850FT) underneath. We first drive the inner piezo actuator from 0 V to 120 V, followed by -50 V, back to 120 V, and ended up at 0 V. While applying voltage on the inner piezo actuator; the outer piezo actuators were connected in a short circuit. At the same time, we measured changes in the resistance of the strain gauge. With the given gauge factor of 1.84 from the manufacturer, the strain was calculated. 
The results are shown in Fig.\ \ref{fig:supFigStrain} (b) in the black line. A clear hysteresis was detected with the maximum compressive strain of $\sim\,-0.42\,\%$ at 120 V and maximum tensile strain of $\sim\,0.27\,\%$ at -50 V. We then put the inner piezo actuator in a short circuit and derive the outer piezo actuators with the same applied voltage sequence. The red line in Fig.\ \ref{fig:supFigStrain} (b) shows a similar hysteresis loop but the opposite sign compared to the hysteresis loop from the inner piezo actuator. The maximum compressive strain of $\sim\,-0.3\,\%$ at -50 V and the maximum tensile strain of $\sim\,-0.39\,\%$ at 120 V were observed.   

\newpage\subsection{Estimate of Defect Density}
For the Te-poor compound \hfd, Lv et al. \cite{Lv2018Tu} have reported a linear relationship between the Te deficit $\delta$ with the temperature of the maximum longitudinal resistivity $\rho_{xx}$, which is plotted in Fig.\ \ref{fig:supFigResistivity} (red dots).  Since this temperature is very similar to $T_{max}$ as defined in our main text, we have used their transport measurements to derive estimates of $\delta$ for our samples (1,2), which are $\delta$ = (.066, .022), respectively.
\supFigResistivity

\newpage\subsection{Strain dependent band structure}
Figure \ref{fig:supFigStrainBS} presents band structure changes with full-cycle piezo actuator movements. (0\,V - compress - tensile - compress) It demonstrates not only apparent changes in band structure with the applied uniaxial stress but also reproducible results.  
\supFigStrainBS

\newpage\subsection{Comparison between ARPES and IGF calculations}
Figures \ref{fig:isoenergy} (a)-(c) and (g)-(i) show constant energy contour plots of \hf\ under compressive strain (sample 2). Corresponding IGF calculations are also plotted in Figs.\ \ref{fig:isoenergy} (d)-(f) and (j)-(l). The binding energies of each plot, Figs.\ \ref{fig:isoenergy} (a)-(l), are marked in Figs.\ \ref{fig:isoenergy} (m) and (n). Comparing ARPES and IGF calculations, we find that sample 2 with compressive strain (or unstrained sample 1) is very close to but not exactly the same as the equilibrium IGF calculation results. First of all, the chemical potential is shifted about 150\,meV downward in ARPES data compared to the IGF calculations. More specifically, the Fermi surface from ARPES measurement is composed of dots which are the residual intensities from the top of the hole bands below $E_{F}$ (Fig.\ \ref{fig:isoenergy} (a)). On the other hand, $E_{F}$ crosses the hole bands in the IGF calculation. As a result, the astroid shape of the Fermi surface is observed in the IGF calculations (Fig.\ \ref{fig:isoenergy} (d)). The astroids become more clear at higher binding energies in both ARPES and IGF calculation results (Fig.\ \ref{fig:isoenergy} (b), (c), and (d)-(f)). ARPES results at the binding energies of 250\,meV and 500\,meV are very similar to the IGF calculation results at 100\,meV and 300\,meV, respectively. This means 150\,meV - 200\,meV band shifts in ARPES data. Secondly, the spacing between two surface bands, $\alpha$, and $\beta$, gets larger with tensile strain in the IGF calculations. (See Fig.\,\ref{fig:band} (h) and (i)) In fact, the spacing between the surface bands in compressed sample 2 and unstrained sample 1 is slightly larger than that of the IGF calculation at equilibrium, which means the samples at those conditions are slightly stretched. 
\supfigisoenergy

\newpage\subsection{Linear band crossings away from high symmetry points}
The astroids shown in Fig.\ \ref{fig:isoenergy} start to cross at \eb\,=\,500\,meV. Crossing points are marked with red dashed lines in Figs.\ \ref{fig:isoenergy}(c), and (g)-(i). With larger binding energy, the points move outwards in the $k_{x}$ direction. Interestingly, the crossing points of the steroids marked in Fig.\ \ref{fig:isoenergy} with red dashed lines correspond to linear crossing in the band structure as shown in Fig.\ \ref{fig:linear}.  
\supfiglinear

\newpage\subsection{Effect of vacancies and strain on bond lengths}

The structural changes are quantified by comparing the Hf-Hf nearest-neighbor distances in the A, B, and C directions as indicated on the crystal structure in Fig. \ref{fig:theory-defects}(a) with the values for the stoichiometric case. For each calculation, one Te vacancy was included in a $2\times 2\times 2$ super cell comprising 16 formula units of \hf\, resulting in a 1.25\% defect density.

\begin{table}[h!]
\caption{\label{tab:bondlengths} Bond lengths in \AA\, as shown in Fig. \ref{fig:theory-defects}(a), for structures with different Te vacancies and strain.} 
\begin{tabular}{l*{3}{>{\centering\arraybackslash}p{4em}}}
\toprule
Structure       & A    & B    & C    \\
\midrule
2\% Compressive & 3.876 & 7.547 & 7.337 \\
\myrowcolour
Equilibrium     & 3.953 & 7.545 & 7.327 \\
2\% Tensile     & 4.034 & 7.554 & 7.311 \\
\myrowcolour
Te-1 Vacancy    & 3.927 & 7.555 & 7.299 \\
Te-2 Vacancy    & 3.956 & 7.568 & 7.322 \\
\myrowcolour
Te-3 Vacancy    & 3.958 & 7.561 & 7.302 \\
\bottomrule
\end{tabular}
\end{table}

\supFIGSTRcal

\newpage

\subsection{Detailed analysis of ARPES data at different strains}

Fig.\ \ref{fig:supFigArpesFit} (b) and (c) indicate a decrease in the amplitude of the bands and an increase of background noise with tensile strain in HfTe$_{5}$.
\supFigArpesFit

\newpage